\documentclass[envcountsame]{llncs}

\usepackage{floatrow}
\usepackage{capt-of}
\usepackage{etex}
\usepackage{amssymb}
\usepackage{amsmath}
\usepackage{booktabs}
\usepackage{cite}
\usepackage{enumerate}
\usepackage{stmaryrd} % provides \bigsqcap
\usepackage{proof}
\usepackage{xspace}
\usepackage{thmtools}
\usepackage{mathtools}
\usepackage{thm-restate}
\usepackage{todonotes}
\usepackage{tikz}
\usetikzlibrary{calc}
\usetikzlibrary{decorations.pathreplacing,calc,decorations.pathmorphing}
\usepackage{tikz-qtree}
\usepackage{url}
\usepackage{graphicx}
\usepackage{comment}

\usepackage{mathrsfs}
\usepackage{multirow}
\usepackage{multicol}
\usepackage{lipsum}
 \usepackage{caption}

\usepackage{multirow}
\usepackage{mathtools}
\usepackage{bm}
\usepackage{verbatim}
\usepackage{extarrows}
\usepackage{setspace}
\usepackage{framed}

\usetikzlibrary{shapes,arrows}
\usepackage{enumitem}

\usepackage[ruled]{algorithm}
\usepackage[noend]{algpseudocode}

\algdef{SE}[DOWHILE]{Do}{doWhile}{\algorithmicdo}[1]{\algorithmicwhile\ #1}
\usepackage{multirow}
\usepackage{mathtools}
\usepackage{bm}
\usepackage{verbatim}
\usepackage{extarrows}
\usepackage{setspace}

\usepackage{booktabs}

\newcommand{\SetEnum}[1]{\{#1\}}

\newcommand{\union}{\cup}

\newcommand{\intersection}{\cap}

\newcommand{\clauseSubsumption}{\ensuremath{\leq_{s}}}

\newcommand{\clauseSetFilteredSubsumption}[1]{\ensuremath{N^{\clauseSubsumption}}}

\newcommand{\EL}{\ensuremath{\mathcal{E\!L}}\xspace}

\newcommand{\ALC}{\ensuremath{{\cal ALC}}\xspace}

\newcommand{\SHIQO}{\ensuremath{{\cal SHOIQ}}\xspace}

\newcommand{\sig}{\ensuremath{\mathsf{sig}}}

\newcommand{\Cmc}{\ensuremath{\mathcal{C}}\xspace}

\newcommand{\Emc}{\ensuremath{\mathcal{E}}\xspace}

\newcommand{\Imc}{\ensuremath{\mathcal{I}}\xspace}
\newcommand{\Jmc}{\ensuremath{\mathcal{J}}\xspace}
\newcommand{\Lmc}{\ensuremath{\mathcal{L}}\xspace}
\newcommand{\Mmc}{\ensuremath{\mathcal{M}}\xspace}

\newcommand{\Omc}{\ensuremath{\mathcal{O}}\xspace}  % for arbitrary ontologies
\newcommand{\Rmc}{\ensuremath{\mathcal{R}}\xspace}  % ALC language extension
\newcommand{\Smc}{\ensuremath{\mathcal{S}}\xspace}
\newcommand{\Tmc}{\ensuremath{\mathcal{T}}\xspace}
\newcommand{\Umc}{\ensuremath{\mathcal{U}}\xspace}  % ALC language extension
\newcommand{\Vmc}{\ensuremath{\mathcal{V}}\xspace}  % ALC language extension

\algnewcommand\algorithmicinput{\textbf{INPUT:}}
\algnewcommand\INPUT{\item[\algorithmicinput]}

\newfloatcommand{capbtabbox}{table}[][\FBwidth]
\newcommand{\wrt}{w.r.t.\xspace}

\newcommand{\Just}{\ensuremath{\mathrm{Just}}\xspace}
\newcommand{\Rep}{\ensuremath{\mathrm{Rep}}\xspace}

\newcommand{\Nodes}{\ensuremath{\textrm{Nodes}}\xspace}

\newcommand{\Mbb}{\ensuremath{\mathbb{M}}\xspace}
\newcommand{\Sbb}{\ensuremath{\mathbb{S}}\xspace}

\newcommand{\Qbb}{\ensuremath{\mathbb{Q}}\xspace}

\newcommand{\Wbb}{\ensuremath{\mathbb{P}}\xspace}

\newcommand{\JustCore}{\textsc{Compute-Justification-Core}\xspace}
\newcommand{\SingleJust}{\textsc{Single-Justification}\xspace}
\newcommand{\JustUnion}{\textsc{Union-of-All-Justifications}\xspace}

\def\thepage{}

\title{Union and Intersection of all Justifications}

\author{
Jieying Chen\inst{1}
\and 
Yue Ma\inst{2}   
\and
Rafael Pe\~naloza\inst{3}
\and
Hui Yang\inst{2}
}
\institute{SIRIUS, Department of Information, University of Oslo, Norway \\ \email{jieyingc@ifi.uio.no}
\and LISN, Univ. Paris-Sud, CNRS, Université Paris-Saclay, Orsay, France
  \\  \email{\{ma, yang\}@lri.fr}
\and University of Milano-Bicocca, Milan, Italy
    \\
   \email{rafael.penaloza@unimib.it}
 }

\begin{document}

\maketitle

\begin{abstract}
% In the paper we present the algorithms for 
% computing core, i.e., the intersection, and union of all justifications \wrt a conclusion, without
% necessarily computing all justifications.
% Additionally, we define a notion, optimal repair, and also show how we could use core and union of all justifications to repair ontologies. 
% Our evaluation shows that, computing the justifications core could be quite efficient even for expressive description logics 
% and our algorithms, especially black-box algorithm, of computing union of all justifications is much faster
% than computing all justifications by state-of-the-art algorithms for expressive description logics.
We present new algorithms for computing the union and intersection of all justifications for a given ontological consequence without first 
computing the set of all justifications. 
Through an empirical evaluation, we show that our approach works well in practice for expressive description logics. 
In particular, the union of all justifications can be computed much faster than with existing justification-enumeration approaches.  
We further discuss how to use these results to repair ontologies.

\end{abstract}

\pagestyle{empty}

% !TEX root =  dl21.tex
\section{Introduction}

A justification for a consequence $\alpha$ refers to a minimal subset of the ontology, which still entails $\alpha$. The problem
of computing justifications, also known as axiom pinpointing, has been widely studied in the context of description 
logics~\cite{Pena-PP20}. Axiom pinpointing methods can be separated into two main classes, commonly known
as \textit{black-box} and \textit{glass-box}.

Black-box approaches~\cite{kalyanpur2005debugging,kalyanpur2006debugging,parsia2005debugging} use existing reasoners as an 
oracle, and require no further modification of the reasoning method.
Therefore, these approaches work for ontologies written in any monotonic logical language (including expressive DLs
such as \SHIQO), as long as a reasoner supporting it exists. In their most na\"ive form, black-box methods check all possible subsets 
of the ontology for the desired entailment and compute the justifications from these results. In reality, many optimisations have been
developed to reduce the number of calls needed, and avoid irrelevant work.

Glass-box approaches, on the other hand, modify the reasoning algorithm to output one or all justifications directly, from only one call.
While the theory for developing glass-box methods has been developed for tableaux and automata-based reasoners 
\cite{BaPe-JLC10,BaPe-JAR10,baader1995embedding,BaPS-KI07}, in practice not many of these methods have been implemented, as 
they require new implementation
efforts and deactivating the optimisation techniques that make reasoners practical.
A promising approach, first proposed in \cite{SeVe-CADE09} is to reduce, through a reasoning simulation, the axiom pinpointing
problem to an enumeration problem from a propositional formula, and use state-of-the-art SAT-solving methods to enumerate 
all the justifications. This idea has led to effective axiom pinpointing systems developed primarily for the lightweight DL \EL
\cite{PULi,beacon,EL2MCS,EL2MUS,SATpin}. 

%On the other hand, \textit{glass-box} approaches go deep into the reasoning process and focus on lightweight description logics \EL as reasoning on $\EL$-ontologies is much simpler and more efficient. 
%There are mainly two different glass-box approaches: 
%one is tableaux-based~\cite{, baader2008automata,baader2006efficient} and
%the other one is SAT-based~
%In general, SAT-based tools, such as PULi~\cite{PULi}, can compute justifications quite efficient on \EL-ontologies in real-world scenario. 
%The SAT-based methods are implied on the light-weight description logic $\mathcal{EL}$ and 
%performs the best because of polynomial-time reasoning over $\mathcal{EL}$ is possible.

% While the \textit{glass-box} approaches have achieved  better performances over certain %a 
% specific description logics  (such as $\mathcal{EL}^+$-ontology) by going deep into the reasoning process. 
% Among them, SAT-based methods \cite{PULi,beacon,EL2MCS,EL2MUS,SATpin} perform the best.

The interest of axiom pinpointing goes beyond enumerating justifications.
Modelling ontologies is a time-consuming and fallible task. Indeed, during the modelling phase it is not uncommon to 
discover unexpected or wrong entailments. One way to fix these errors is to diagnose the causes by computing a hitting
set of all the justifications.
However,  as there might exist exponential many justifications for a given entailment w.r.t.\ an ontology,
even for \EL-ontologies,
finding all justifications is not feasible in general.
One approach is to approximate the information by the union and intersection of all justifications. If the intersection is not
empty, then any axiom in this intersection, when removed, guarantees that the consequence will not follow anymore. From the
union, a knowledge engineer has a more precise view on the problematic instances, and can make a detailed analysis.

Although much work has focused on methods for computing one or all justifications efficiently, to the best of our knowledge
there is little work on computing their intersection or union without enumerating them first, beyond the approximations presented
in \cite{conf/esws/PenalozaMIM17,journals/ki/Penaloza20}.
In this paper, we propose an algorithm of computing the intersection of all justifications.
This algorithm has the same worst-case behaviour as the black-box algorithm of computing one justification.
Additionally, we present two approaches of computing the union of all justifications, one is based on the black-box algorithm of finding 
all justifications and the other approach uses the SAT-tool \textit{cmMUS}.

The paper is structured as follows. 
In Section~\ref{sec:preliminaries} we recall relevant definitions of description logics and propositional logic.
Section~\ref{sec:core} presents the algorithm for computing the intersection of all justifications without computing any single justification.
We propose two methods of computing the union of all justifications in Section~\ref{sec:union}.
We explain how to use the union and intersection of all justifications to repair ontologies in Section~\ref{sec:application}. 
Before concluding, an evaluation of our methods on real-world ontologies is presented in Section~\ref{sec:evaluation}.

% !TEX root =  dl21.tex

\section{Justifications and Repairs in \ALC}
\label{sec:preliminaries}

We briefly recall the notions of justifications and repairs in \ALC.
Let $N_C$, $N_R$ and $N_I$ be mutually disjoint sets of \emph{concept names}, \emph{role names}, and \emph{individual names}. 
The set of \ALC-concepts $C$ 
is built through the following grammar rule
\begin{align*}
    C &::= \top \mid \bot \mid A \mid C\sqcap C \mid C\sqcup C \mid \neg C \mid \exists r. C \mid \forall r. C, 
\end{align*}
where $A\in N_C$ and $r\in N_R$.
An \ALC-TBox $\Tmc$ is a finite set of general concept inclusions (GCIs) of the form $C\sqsubseteq D$ and role 
inclusions $r\sqsubseteq s$, where $C$ and $D$ are \ALC-concepts and $r,s \in N_R$.
An ABox is a finite set of concept assertions of the form $A(a)$ and role assertions $r(a,b)$, 
where $A\in N_C$, $r\in N_R$ and $a,b\in N_I$.
An \ALC ontology consists of an \ALC TBox and an ABox.

The semantics of this logic is defined in terms of interpretations. 
An interpretation is a pair
$\mathcal{I} = (\Delta^\mathcal{I},\cdot^\mathcal{I})$ where $\Delta^\Imc$ is a non-empty set called the \emph{domain}, and
$\cdot^\Imc$ is the \emph{interpretation function}, which
maps each concept name $A \in N_C$ to a subset $A^\mathcal{I} \subseteq \Delta^\mathcal{I}$,
each role name $r \in N_R$ to a binary relation 
$r^\mathcal{I} \subseteq \Delta^\mathcal{I}{\times}\Delta^\mathcal{I}$
 and each individual $a\in N_I$ to a domain element $a^\Imc \in \Delta^\Imc$.
The interpretation function is extended to \ALC-concepts as usual:
% \begin{itemize}
%     \item $(\top)^\mathcal{I} {=} \triangle^\mathcal{I}$,
%     \item $( C\sqcap D)^\mathcal{I} {=} C^\mathcal{I}{\cap} D^\mathcal{I}$,
%     \item $( C\sqcup D)^\mathcal{I} {=} C^\mathcal{I}{\cup} D^\mathcal{I}$,
%     \item $(\neg C)^\mathcal{I} {=} \triangle^\mathcal{I}\setminus C^\mathcal{I}$,
%     \item $ (\exists r. C)^\mathcal{I} {=} \{a{\in} \triangle^\mathcal{I} \mid \exists b{\in} C^\mathcal{I}, (a,b){\in} r^\mathcal{I}\}$;
%     \item $ (\forall r. C)^\mathcal{I} {=} \{a{\in} \triangle^\mathcal{I} \mid \text{ if }(a,b){\in} r^\mathcal{I},\text{ then }b{\in} C^\mathcal{I}\}$.
% \end{itemize}
$(\top)^\Imc := \Delta^\Imc$,
$(\bot)^\Imc := \emptyset$,
$(\neg C)^\Imc :=  \Delta^\Imc \backslash C^\Imc$,
$(C\sqcap D)^\Imc := C^\Imc \cap D^\Imc$, 
$(C\sqcup D)^\Imc := C^\Imc \cup D^\Imc$, 
$(\exists r.C)^\Imc := \{ x\in \Delta^\Imc \mid \exists y\in C^\Imc : (x,y)\in r^\Imc\}$, and
$(\forall r.C)^\Imc := \{ x\in \Delta^\Imc \mid \forall y\in\Delta^\Imc. (x,y)\in r^\Imc \Rightarrow y\in C^\Imc \}$.
The interpretation \Imc \emph{satisfies} $C\sqsubseteq D$
iff $C^\Imc\subseteq D^\Imc$ and it \emph{satisfies} $r\sqsubseteq s$ %($r\sqsubseteq s$) 
iff $r^\Imc\subseteq s^\Imc$.
We write $\Imc\models \alpha$ if \Imc satisfies the axiom $\alpha$. 
The interpretation \Imc is a \emph{model} of an ontology \Omc if \Imc satisfies all axioms in \Omc.
An axiom $\alpha$ \emph{is entailed by \Omc}, denoted as $\Omc \models \alpha$,
if  $\Imc\models\alpha$ for all models \Imc of \Omc.
%
% If for all $C{\sqsubseteq} D, r{\sqsubseteq} s{\in} \mathcal{O}$,  we have $C^\mathcal{I}{\subseteq} D^\mathcal{I}, r^\mathcal{I}{\subseteq} s^\mathcal{I}$ respectively,
% we say the interpretation $\mathcal{I}$ is compatible with all axioms in $\Tmc$ and call $\mathcal{I}$ a \textit{model} of $\Tmc$. Given a \Lmc-axioms $\alpha$, 
% we say $\Tmc{\models} a$ iff each model of $\Tmc$ is compatible with  $\alpha$.
%
We use $|\Omc|$ to denote the size of $\Omc$, i.e., the number of axioms in $\Omc$.

For this paper, we are interested in the notions of \emph{justification} and \emph{repair}.

\begin{definition}[Justification, repair]
\label{def:just}
Let $\Omc$ be an ontology and $\alpha$ a GCI.
A \emph{justification} for $\Omc \models \alpha$ is a subset $\Mmc \subseteq \Omc$
such that $\Mmc \models \alpha$ and for any $\Mmc' \subsetneq \Mmc$,
$\Mmc' \not\models \alpha$. 
$\Just(\Omc, \alpha)$ denotes the set of all justifications of $\alpha$ w.r.t.\ \Omc.
%\end{definition}
%
%\begin{definition}[Repair]
%Let $\Omc$ be an ontology and let $\alpha$ be a conclusion of the form $C \sqsubseteq D$.
A repair for  $\Omc \models \alpha$ is a subontology $\Rmc \subseteq \Omc$ such that 
$\Rmc \not\models \alpha$, but $\Omc' \models \alpha$ for any $\Rmc \subsetneq \Omc'\subseteq \Omc$.
We denote the set of all repairs as $\Rep(\Omc,\alpha)$.
\end{definition}
Briefly, a justification is a minimal subset of an ontology that preserves the conclusion.
Dually, a repair is a maximal sub-ontology that does not preserve the consequence.

%+IAR semantic

Now we consider a propositional language  
%In propositional logic, given
with a finite set of propositional variables
%a literal set 
$L = \{p_1, p_2, \cdots, p_n\}$. A literal is a variable $p_i$ or its negation $\neg p_i$. A clause $l_1 \vee l_2\vee \cdots \vee l_k$ is a disjunction of literals,   
%a disjunction of literals and their negation $\neg p_{i_1}\vee \cdots \vee \neg p_{i_k}\vee p_{j_1}\vee \cdots \vee p_{j_t}$ is called a \textit{clause} and is 
denoted
by $\omega$\cite{chang2014symbolic}.
A Boolean formula in Conjunctive Normal Form (CNF) 
is a conjunction of clauses.  
A CNF formula $\phi$ is \textit{satisfiable} iff there exists a \textit{truth assignment} $\mu_L: L\rightarrow \{0,1\}$ such that $\mu_L$ 
satisfies all clauses in $\phi$. 
We can also consider a CNF formula as a set of clauses. 
A subformula $\phi'\subseteq \phi$ is a \textit{Minimally Unsatisfiable Subformula (MUS)} iff $\phi'$ is unsatisfiable, but for every 
$\phi_1'\subsetneq \phi'$ is satisfiable.

% !TEX root =  dl21.tex

\section{Computing the Intersection of all Justifications}\label{sec:core}

We first study the problem of computing the intersection of all justifications, which we often call the \emph{core}. 
Algorithm~\ref{alg:compute_core} provides a method for finding this core.
\begin{algorithm}[tb]
\caption{\small  Computing the intersection of all justifications of $\Omc$ \wrt $\alpha$}
\label{alg:compute_core}
%\setstretch{1.2}
\small
\begin{algorithmic}[1]
\INPUT an Ontology $\Omc$, a conclusion $\alpha$
\Function{$\JustCore$}{$\Omc, \Sigma$}
%     \State $\Omc_\Sigma \coloneqq \textsc{SyntacticLocalityModule}$($\Omc, \Sigma$)
    \State $\Cmc \coloneqq \emptyset$
    \State $\Mmc := \textsc{Compute-Locality-Based-Module}(\Omc,\sig(\alpha))$
    \For{every axiom $\beta \in \Mmc$}
            \If{$ \Mmc \setminus \SetEnum{\beta} \not\models \alpha$}
                \State $\Cmc \coloneqq \Cmc \union \SetEnum{\beta}$
            \EndIf
    \EndFor
    \State \Return $\Cmc$
    \EndFunction
\end{algorithmic}
\end{algorithm}
The algorithm is %{based on}
{inspired by} the known black-box approach for finding justifications \cite{KPHS07,BaPS-KI07}. Starting from a
justification-preserving module \Mmc (in this case, the locality-based module, Line 3), we try to remove one axiom $\beta$ (Line 4). 
If the removal of the axiom 
$\beta$ removes the entailment (Line 5), then
$\beta$ must belong to all justifications ($\beta$ is a \emph{sine qua non} requirement for entailment within \Mmc), and is thus added
to the core \Cmc (Line 6).

Algorithm~\ref{alg:compute_single_just}, on the other hand, generalises the known algorithm for computing a single justification,
by considering a (fixed) set \Cmc that is known to be contained in all justifications. If $\Cmc=\emptyset$, the approach works
as usual; otherwise, the algorithm avoids trying to remove any axiom from \Cmc. This reduces the number of calls to the
black-box reasoner, potentially decreasing the overall execution time.
%This can reduce the overall execution time 
%of the black-box approach, for which the core is pre-computed merely once but computing single justifications can be executed exponentially. \todo{check this revised sentence}
%
\begin{algorithm}[tb]
\caption{\small Using core to compute a single justification of $\Omc$ $\wrt$ an conclusion %$\alpha$
}
\label{alg:compute_single_just}
%\setstretch{1.2}
\small
\begin{algorithmic}[1]
\INPUT an ontology $\Omc$, a conclusion $\alpha$, the intersection of all justifications $\Cmc$

\Function{$\SingleJust$}{$\Omc, \alpha, \Cmc$}
%     \State $\Omc_\Sigma \coloneqq \textsc{SyntacticLocalityModule}$($\Omc, \Sigma$)
    \State $\Mmc \coloneqq \textsc{Compute-Locality-Based-Module}(\Omc,\sig(\alpha))$
    \For{every axiom $\beta \in \Mmc$ and $\beta \not\in \Cmc$}
            \If{$ \Mmc \setminus \SetEnum{\beta} \models \alpha$}
                \State $\Mmc \coloneqq \Mmc \setminus \SetEnum{\beta}$
            \EndIf
    \EndFor
    \State \Return $\Mmc$
    \EndFunction
\end{algorithmic}
\end{algorithm}
%
%Note that Algorithms \ref{alg:compute_core} and \ref{alg:compute_single_just} can be executed together to find the first justification
%and the core simultaneously.

As mentioned already, the choice for a locality-based module in these algorithms is arbitrary, and any justification-preserving module
would suffice. In particular, we could compute lean kernel~\cite{conf/esws/PenalozaMIM17,conf/ijcai/KoopmannC20} for \ALC-ontologies, 
and minimal subsumption modules~\cite{ChenLMW-ISWC17,conf/gcai/ChenL018} for \EL-ontologies
instead, which is typically smaller thus reducing the number of iterations within the algorithms.
However, as it could be quite expensive to compute such modules,
it might not be worthwhile in some cases. 
The following theorem shows that Algorithm~\ref{alg:compute_core} correctly computes the intersection of all justifications.

%\begin{restatable}{theorem}{ThmJustCoreCorr}
\begin{theorem}
Let $\Omc$ be an ontology and 
$\alpha$ a GCI.
Algorithm~\ref{alg:compute_core} %applied on~$\Omc$ and~$\alpha$
computes the intersection of all justifications of~$\Omc$ \wrt~$\Sigma$.
\end{theorem}
%\end{restatable}
%
Algorithm~\ref{alg:compute_core}, like all black-box methods for computing justifications,
calls a standard reasoner $|\Mmc|$ times. In terms of computational complexity, computing the core requires as many
computational resources as computing a single justification.
However, computing one justification might be faster in practice, as the size of \Mmc decreases throughout the execution of
Algorithm~\ref{alg:compute_single_just}.
Clearly, if the core coincides with one justification \Mmc, then \Mmc is the \emph{only} justification.
\begin{corollary}
\label{cor:int:one}
Let $\Omc$ be an ontology, $\alpha$ a GCI; and let $\Cmc$ be the core and \Jmc a justification for $\Omc \models \alpha$.
If $\Cmc=\Jmc$, \Jmc is the only justification for $\Omc \models \alpha$.
\end{corollary}

% !TEX root =  dl21.tex

\section{Computing the Union of all Justifications}
\label{sec:union}

We now present two algorithms of computing the union of all justifications. 
The first algorithm follows a black-box approach that calls a standard reasoner as oracle using the core of justifications. This is inspired by Reiter's Hitting Set Tree
algorithm~\cite{ReiterDiagnosis} and partially in line with \cite{KPHS07,10.1007/978-3-540-89704-0_1}. 
For the second algorithm, we reduce the problem of computing the union of all justifications to the problem of computing the union of 
MUSes of a propositional formula.
Note that the second algorithm works only for \ALC-ontologies, while
the first algorithm can be applied to ontologies with any expressivity, as long as a reasoner is available.

\subsection{Black-box algorithm}

The black-box algorithm of computing all justifications~\cite{10.1007/978-3-540-89704-0_1} was inspired by the algorithm of 
computing all minimal hitting sets~\cite{ReiterDiagnosis}.
Some of the improvements to prune the search space were already proposed in~\cite{ReiterDiagnosis}.
Our method for computing the union of all justifications (Algorithm~\ref{alg:compute_union}) works in a similar manner, but with 
a few key differences.
%
%\todo{what is the initialization of $\rho$?}
\begin{algorithm}[tb]
\caption{Computing the Union of All Justifications \wrt a Conclusion $\alpha$}
\label{alg:compute_union}
%\setstretch{1.2}
\small
\begin{algorithmic}[1]
\INPUT
an Ontology $\Omc$, a conclusion $\alpha$,
        the intersection of all Justifications $\Cmc \subseteq \Omc$ 
\Function{$\textsc{Union-of-All-Justifications}$}{$\Omc, \alpha, \Cmc$}
        \State $\Mmc \coloneqq \textsc{Compute-Locality-Based-Module}$($\Omc, \sig(\alpha))$
        \State $\Umc \coloneqq \Cmc$;
        $\Psi \coloneqq (\SetEnum{\rho}, \emptyset, \emptyset, \rho)$;
        $\Qbb \coloneqq [\rho]$; 
         $\Wbb \coloneqq \emptyset$;
         $\Mbb \coloneqq \{\emptyset\}$
        \While{$\Qbb \neq [\,]$}
            \State $v \coloneqq \textsc{Head}(\Qbb)$, $\Qbb \coloneqq \textsc{RemoveFirstElement}(\Qbb)$, $\Wbb \coloneqq \Wbb \union \SetEnum{v}$
            \State $\Mmc_\text{ex} \coloneqq\textsc{Labels}(\textsc{Path}(\Psi, \rho, v))$
%             \State $\Obb \coloneqq \Obb \union \SetEnum{\Omc_\text{ex}}$
            \If{
            $\textsc{Is-Path-Redundant}(\Psi,\rho,\Mmc_{\text{ex}},\Wbb)$
%             there exists $\pi' \in \textsc{Paths}(\Psi, \rho)$ such that
%             $\pi_v \neq \pi'$ and
%             $\Omc_\text{ex} = \textsc{Labels}(\pi')$
            }
                    \State \textbf{continue}
            \EndIf
            \If{$ \Mmc \setminus \Mmc_\text{ex} \not\models \alpha$}
                \State \textbf{continue}
            \EndIf
            \If{$\Mmc \setminus \Mmc_\text{ex} \subseteq \Umc$}
                \State \textbf{continue}
            \EndIf
            \State $\Mmc \coloneqq \emptyset$
            \If{there exists $\Mmc' \in \Mbb$ such that $\Mmc_\text{ex} \intersection \Mmc' =\emptyset$}
                    \State $\Mmc \coloneqq \Mmc'$
            \Else
                \State $\Mmc \coloneqq \textsc{Single-Justification}(\Mmc \setminus \Mmc_\text{ex}, \alpha,  \Cmc)$
                \If{$\Mmc = \Cmc$}
			\State \Return $\SetEnum{\Cmc}$
                \EndIf
                \State $\Mbb \coloneqq \Mbb \union \SetEnum{\Mmc}$
                \State $\Umc \coloneqq \Umc \cup \Mmc$
 	    \EndIf
            \For{every $\beta \in \Mmc \setminus \Cmc$}
                    \State $v_\beta \coloneqq \textsc{AddChild}(\Psi, v, \beta)$
                    \State $\Qbb \coloneqq v_\beta :: \Qbb$
            \EndFor
        \EndWhile

	\State \Return $\Umc$
	\EndFunction
\end{algorithmic}
\end{algorithm}
To avoid computing all justifications, we prune the search space when
all remaining justifications are fully contained in the union computed so far (Lines~11-12). 
In addition, we use the core to speed the search. 
As the axioms in the core must appear in every justification, we can reduce the number of calls made to the 
reasoner, and optimise the single justification computation (Line 17).
Finally, when we organise our search space, we do not need to consider the axioms in the core (Line 22).

We now describe the $\textsc{Union-of-All-Justifications}$ procedure in detail.
Given an ontology $\Omc$,
a signature $\Sigma$,
and the intersection of all justifications $\Cmc \subseteq \Omc$ of~$\Omc$ \wrt~$\Sigma$ as input,
a syntactic $\bot\!\top^\ast$-locality module~$\Omc_\Sigma$ of~\Omc \wrt~$\Sigma$ is extracted
from~\Omc (Lines~2).
The justification search tree $\Psi$ is a four-tuple $(\Vmc, \Emc, \Lmc, \rho)$, where $\Vmc$ is a finite
set of nodes,
$\Emc \subseteq \Vmc \times \Vmc$ is a set of edges,
$\Lmc$ is an edge labelling function, mapping every edge  to an axiom~$\alpha \in \Mmc$,
and $\rho \in \Vmc$ is the root node.
We initialise the variable~$\Psi$ to represent a justification search tree for~\Omc having
only root node~$\rho$.
Besides,
the variables $\Mbb \subseteq 2^{\Omc_\Sigma}$, containing the justifications
that have been computed so far,
and $\Wbb \subseteq \Vmc$,
containing the already explored nodes of~$\Psi$,
are both initialised with the empty set.
The queue~$\Qbb$ of nodes in~$\Psi$ that still has to be explored
is also set to contain the node~$\rho$ as its only element.

The algorithm then enters a loop (Lines~4--24)
that runs while~$\Qbb$ is not  empty.
The loop extracts the first element~$v$ from~$\Qbb$  
and adds it to~$\Wbb$ (Line~5).
The axioms that label the edges of the path~$\pi_v$ from~$\rho$ to~$v$ in~$\Psi$
are collected in the set $\Mmc_\text{ex}$ (Line~7).
After that, the algorithm checks whether $\pi_v$ is redundant. 
The detailed method for checking redundancy is described in Algorithm~\ref{alg:is_path_redundant}.
The path $\pi_v$ is \emph{redundant} iff there exists an explored node $w\in\Wbb$ such
that~(a) the axioms in $\Omc_\text{ex}$ are exactly the axioms labelling the edges of the
path~$\pi_w$ from~$\rho$ to~$w$ in~$\Psi$ (Lines 4--6), or (b)~$w$ is a leaf node of~$\Psi$ and the edges
of~$\pi_w$ are only labelled with axioms from~$\Omc_\text{ex}$ (Lines 7--8).
Case~(a) corresponds to \emph{early path termination} in~\cite{ReiterDiagnosis, KPHS07}:
the existence of~$\pi_w$ implies that all possible extensions
of~$\pi_v$ have already been considered.
Case~(b) implies that the axioms labelling the edges of~$\pi_w$ lead to the fact that
$\alpha$ can not be entailed be the remaining TBox
when removed from~$\Omc_\Sigma$.
Therefore, 
by monotonicity of~$\models$, 
we infer that removing~$\Omc_\text{ex}$ from~$\Omc_\Sigma$ also has the same consequence
implying that we do not need to explore $\pi_v$ and all its extensions.
\begin{algorithm}[tb]
\caption{Checking the Redundancy of a Path}
\label{alg:is_path_redundant}
%\setstretch{1.2}
\small
\begin{algorithmic}[1]
\INPUT
justification search tree $\Psi$ for an \Lmc-TBox~\Omc with root $\rho$,
$\Omc^w_\text{ex} \subseteq \Omc$,
$\Wbb \subseteq \Nodes(\Psi)$
\Function{$\textsc{Is-Path-Redundant}$}{$\Psi, \rho, \Omc_\text{ex}, \Wbb$}
	      \For{\textbf{every } $w \in \Wbb$}
			\State $\Omc^w_\text{ex} \coloneqq \textsc{Labels}(\textsc{Path}(\Psi, \rho, w))$
    		\If{$\Omc^w_\text{ex} \subseteq \Omc_\text{ex}$}
    			\If{$\Omc_\text{ex} \subseteq \Omc^w_\text{ex}$}
    				\State \Return \textbf{true}
    			\ElsIf{$\textsc{Is-Leaf}(\Psi, w)$}
    				\State \Return \textbf{true}
    			\EndIf
    		\EndIf
	      \EndFor
		\State \Return \textbf{false}
	\EndFunction
\end{algorithmic}
\end{algorithm}

The current iteration can be terminated immediately if
$ \Mmc \setminus \Mmc_\text{ex} \not\models \alpha$ (Lines~9--10)
as no subset of~$\Mmc_\Sigma \setminus \Mmc_\text{ex}$ can be a justification
of~$\Mmc$ \wrt~$\alpha$.
In contrast to other black-box algorithms for computing justifications, 
we additionally check whether $\Mmc \setminus \Mmc_\text{ex}$ is
a subset of $\Umc$.
If so, no new axioms belonging to the union of all justifications appear in this sub-tree.
Hence, the algorithm does not need to explore it any further. 
Subsequently, the variable~$\Mmc$ that will hold a justification of~$\Mmc \setminus \Mmc_\text{ex}$ is initialised with~$\emptyset$.
At this point we can check if a justification~$\Mmc' \in \Mbb$ has already been computed
for which $\Omc_\text{ex} \intersection \Mmc' = \emptyset$ (Lines~14--15) holds,
%\todo{for line 14, since $\emptyset\in\Mbb$always true  by Lines 3 and 20, I think this condition also holds always; no? Line 3 might be modified...}
in which case%\todo{. If it is the case we set...} 
we set~$\Mmc$ to~$\Mmc'$.
This optimisation step can also be found in~\cite{ReiterDiagnosis,KPHS07} and it allows us
to avoid a costly call to the \textsc{Single-Justification} procedure.
Otherwise, in Line~17 we call \textsc{Single-Justification} on~$\Mmc \setminus \Omc_\text{ex}$
to obtain a justification of $\alpha$ w.r.t.\ $\Mmc \setminus \Omc_\text{ex}$.
We then check whether
$\Mmc$ is equal to $\Cmc$ (Lines~18--19), in which case the search for
additional justifications can be terminated (recall Corollary \ref{cor:int:one}).
Otherwise, the justification~$\Mmc$ is added to~$\Mbb$ in Line~20 and the union of all justifications is updated in Line~21.
Finally, for every $\beta \in \Mmc \setminus \Cmc$,
the algorithm extends the tree $\Psi$ in Lines~22--24 by adding a child $v_\alpha$ to~$v$,
connected by an edge labelled with~$\beta$.
Note that it is sufficient to take $\beta \not \in \Cmc$ as a set~$\Mmc$ with
$\Cmc \not \subseteq \Mmc$ cannot be a justification of~$\Omc$ \wrt~$\alpha$.
The procedure finishes by returning the set~$\Umc$.

Note that this algorithm only adds justifications to~$\Mbb$.
For completeness, one can show that the locality-based module~$\Omc_\Sigma$
of~$\Omc$ \wrt~$\Sigma$ contains all the minimal modules of~$\Omc$ \wrt~$\Sigma$.
Moreover, it is easy to see that the proposed optimisations do not lead to a minimal
module not being computed.
Overall, we obtain the following result.

%\begin{restatable}{theorem}{ThmUnionCorr}
\begin{theorem}
Let \Omc be an ontology, $\alpha$ a GCI, and
$\Cmc \subseteq \Omc$ the core of $\alpha$ w.r.t.\ $\Omc$.
The procedure~$\textsc{Union-of-All-Justifications}$ computes the union of all justifications of~$\Omc$ \wrt~$\alpha$.
\end{theorem}
%\end{restatable}
%
Algorithm~\ref{alg:compute_union} terminates on any input as
the paths in the module search tree~$\Psi$ for~\Omc that is constructed during the
execution represent all the permutations
of the axioms in~\Omc that are relevant for finding all minimal modules.
% The length of any path in the module search tree~$\Psi$ for~\Omc is bounded by $|\Omc|$
%
%
% as either an empty minimal module can be found or the exclusions sets~$\Omc_\text{ex}$
% are extended until no module of~$\Omc_\Sigma \setminus \Omc_\text{ex}$ \wrt~$\Sigma$ can be found.
It is easy to see that the procedure~\JustUnion  runs in
exponential time in size of~$\Omc$ (and polynomially in~$\Sigma$, $n$, and~$\Cmc$)
in the worst case.
%\todo{polynomially in~$\Sigma$, $n$, and~$\Cmc$?? what is $n$? if it is polynomial in $\Cmc$, how it can be exp in $\Omc$?}

% \subsection{Encoding of QBF formula}

% \begin{lemma}
% \label{lemma:axiom_necessary}
% Let $\Omc$ be an \Lmc-TBox, 
% $\alpha$ be a conclusion of the form $C \sqsubseteq D$, and $\beta \in\Omc$.
% Additionally, let $\Umc$ be the union of all justifications of $\Omc$ \wrt $\alpha$.
% Then 
% $\beta\in\Umc$ iff there exists an $\Mmc \subseteq \Omc$ such that $\beta \in \Mmc$,  $\Mmc \models \alpha$, but $\Mmc \setminus \{\beta\} \not\models \alpha$.
% \end{lemma}

% Lemma~\ref{lemma:axiom_necessary} can be used to check whether an axiom from the TBox
% belongs to the union of all justifications.

\subsection{MUS Membership Problem}

We now show how to compute the union of all justifications of a GCI $\alpha$ by a membership approach. The idea is to check 
the membership of each axiom, i.e., whether it is a member of some justification. The main procedure is: firstly, as a pre-processing 
step, we compute a CNF formula $\phi$ using the consequence-based reasoner \textit{condor}%
\footnote{We restrict to $\mathcal{ALC}$ in this section as \textit{condor} only accepts $\mathcal{ALC}$-TBoxes.} 
proposed in \cite{condor}. Then, we compute the union of all justifications of $\alpha\in \Omc$ by checking the membership for 
each axiom using the SAT-tool~\textit{cmMUS}~\cite{janota2011cmmus} and $\phi$. 
In generally, the classification of an $\mathcal{ALC}$-TBox is of exponential complexity. Since the MUS-membership problem is 
$\Sigma_2^P$-complete~\cite{liberatore2005redundancy}, it follows that this method runs in exponential time.

Specifically, the method is divided in two steps:
\begin{enumerate}
    \item \textbf{Compute CNF formula $\phi$.}
    Let $H, K$ denote (possibly empty) conjunctions of concepts, and $M, N$ (possibly empty) disjunctions of concepts;
    	 \textit{condor} classifies the TBox through the inference rules in Table \ref{inference rules}.  
  \begin{table}[t]
    \caption{Inference rules of \textit{condor}}
    \label{inference rules}
    \centering
  \begin{align*}
    &\mathbf{R^+_A}~ \frac{}{H\sqsubseteq A}: A\sqsubseteq H~~~~~~~~~~~~~\mathbf{R^-_A}~ \frac{H\sqsubseteq N\sqcup A}{H\sqsubseteq N}: \neg A\sqsubseteq H\\[0.7mm]
 &\mathbf{R^+_A}~ \frac{\{H\sqsubseteq N_i\sqcup A_i\}_{i=1}^n}{H\sqsubseteq \sqcup_{i=1}^n N_i\sqcup M} : \sqcap_{i=1}^n A_i\sqsubseteq M\in \Omc\\[0.7mm]
 &\mathbf{R^n_\sqcap}~ \frac{H\sqsubseteq N\sqcup A}{H\sqsubseteq N\sqcup \exists R.B} :  A\sqsubseteq \exists R.B\in \Omc\\[0.7mm]
 &\mathbf{R^+_\exists}~ \frac{H\sqsubseteq M\sqcup \exists R.K, K\sqsubseteq N\sqcup A}{H\sqsubseteq M\sqcup  B\sqcup \exists R(K\sqcap \neg A)} : \exists R.A \sqsubseteq B\in \Omc, \Omc\models R\sqsubseteq S\\[0.7mm]
 &\mathbf{R^\perp_\exists}~ \frac{H\sqsubseteq M\sqcup \exists R.K, K\sqsubseteq \perp}{H\sqsubseteq M}\\[0.7mm]
 &\mathbf{R_\forall}~ \frac{H\sqsubseteq M\sqcup \exists R.K, K\sqsubseteq N\sqcup A}{H\sqsubseteq M\sqcup  B\sqcup \exists R(K\sqcap B)} : A \sqsubseteq \forall S.B\in \Omc, \Omc\models R\sqsubseteq S\\
    \end{align*}
\end{table}
Each inference rule can be rewritten as a clause. For example, the $R^+_\exists$ can be transferred to 
$\neg p_1\vee \neg p_2 \vee p_3$ if we denote the 
$H{\sqsubseteq} N{\sqcup} A,\ A{\sqcup} \exists R.B,\  H{\sqsubseteq} N{\sqcup} \exists R.B$ as literals $p_1, p_2, p_3$.
Then the CNF formula $\phi$ is the conjunction of all the clauses corresponding to all the applied inference rules during the classification 
process. For details see \cite{conf/esws/PenalozaMIM17,SeVe-CADE09}. 

    \item\textbf{Check membership of each axiom using \textit{cmMUS}.}
    Given an CNF formula $\phi$ and a subformula $\phi'\subseteq \phi$, the algorithm $\textit{cmMUS}$ is used to determine 
    	whether there is a MUS $\phi''\subseteq \phi$ such that $\phi'\cap \phi''\neq \emptyset$.  We say $\textit{cmMUS}(\phi, \phi') = 1$ 
	if there exists such MUS $\phi''$ and $0$ otherwise. The membership is checked as follows:
    \begin{enumerate}
        \item Define a CNF-formula $\phi_\Omc = \wedge_{\beta\in \Omc} p_\beta$, where each literal $p_\beta$ corresponds to an 
        	axiom $\beta\in \Omc$, and $\phi_\alpha = \neg p_\alpha$, where $\alpha$ is the given conclusion.
        \item Define $\psi_\alpha = \phi\wedge \phi_\Omc\wedge \phi_\alpha$. Then $\psi_\alpha$ is unsatisfiable; each MUS 
        	$\psi'\subseteq \psi_\alpha$ corresponds to a justification of $\alpha$; and $\forall \beta \in \Omc$, 
	$\textit{cmMUS}(\psi_\alpha, p_\beta)=1$ iff $\beta$ belongs to some justifications of $\alpha$. 
    \end{enumerate}
\end{enumerate}
%
%\begin{remark}
Note that only a small number of clauses in $\phi$ are related to the derivation of $\alpha$. In practice,
(i) $\phi'\subseteq \phi$ is the subformula contributing to the derivation of $\alpha$ obtained by tracing back from $\alpha$, 
(ii) $\phi_\Omc'\subseteq \phi_\Omc$ is the subformula including only $\beta\in \Omc$ that appears in $\phi'$. Using 
$\psi_\alpha' = \phi'\wedge \phi_\Omc'\wedge \phi_\alpha$ instead of $\psi_\alpha$ as the input of algorithm \textit{cmMUS} %where $\phi'\subseteq \phi$ is the subformula contribute to the derivation of $\alpha$ obtained by tracing back from $\alpha$, and $\phi_\Omc'\subseteq \phi_\Omc$ is the subformula including only the $\beta\in \Omc$ that appear in $\phi'$. 
%With those simple and fast operations, we 
can significantly accelerate the \textit{cmMUS} algorithm.
%\end{remark}

% !TEX root =  dl21.tex

\section{Repairing Ontologies}
\label{sec:application}

In this section we propose a notion of optimal repair and provide a method for computing all such optimal repairs.

%+ reference of close definition of optimal repairs in inconsistency. 

% delete the union of justifications, 
% ...

\begin{definition}[Optimal Repair]
\label{def:optimalRepair}
Let $\Omc$ be an ontology, $\alpha$ a GCI, and $\Rep(\Omc,\alpha)$ the set of all repairs for $\Omc \models \alpha$.
We say $\Rmc \in \Rep(\Omc,\alpha)$ is an \emph{optimal repair} for $\Omc \models \alpha$, 
if $|\Rmc|\geq|\Rmc'|$ holds for every $\Rmc' \in \Rep(\Omc,\alpha)$.
\end{definition}
That is, an optimal repair is a repair such that removes the least amount of axioms from the original ontology.
It is also important to recall the notion of a hitting set
\begin{definition}[HS]
We say $\Smc$ is a minimal 
\emph{hitting set} for a sets $\Sbb$  if 
%$\Smc$ is an  minimal set \wrt. set inclusion satisfying 
$\Smc \cap s \neq \emptyset$ for every
$s\in\Sbb$. 

\end{definition}
We say $\Smc$ is the smallest minimal hitting set if $|s|$ is the smallest among all minimal hitting set.
The following proposition shows how we can compute the set of all optimal repairs through a hitting set computation
\cite{ScCo-IJCAI03,LiSa-SAT05,BaPe-JLC10}.

\begin{proposition}
\label{prop:HS}
%Let $\Omc$ be an ontology, $\alpha$ be a conclusion and
Let $\Just(\Omc, \alpha)$ be the set of all justifications for the GCI $\alpha$ w.r.t.\ the ontology \Omc.
If $\Sbb$ is the set of all smallest minimal 
hitting sets for $\Just(\Omc, \alpha)$,
then $\{\Omc \setminus \Smc \mid \Smc \in\Sbb\}$ is the set of all optimal repairs for  $\Omc \models \alpha$.
\end{proposition}
%
%+complexity result of computing smallest minimal hitting set
%
When the core is not empty,
a set that consists of only one axiom from the intersection of all justifications is a smallest %minimal 
hitting set for all justifications. 
We get the following corollary, stating how to compute all optimal repairs faster in this case, as a simple consequence of
Proposition \ref{prop:HS}.

\begin{corollary}
\label{cor:optimalRepairAlg}
Let $\Omc$ be an ontology, $\alpha$ a GCI and \Cmc the core for $\Omc \models \alpha$.
If $\Cmc \neq \emptyset$, then $\{\Omc \backslash \{\beta\} \mid \beta \in \Cmc\}$ is the set of all optimal repairs  for  $\Omc \models \alpha$.
% \begin{itemize}
%     \item $\Omc \backslash \{\beta\}$ is an optimal repair for  $\Omc \models \alpha$;
%     \item $\{\Omc \backslash \{\beta\} \mid \beta \in \Cmc\}$ is the set of all optimal repairs.
% \end{itemize}
\end{corollary}
%
%This corollary is an easy consequence of Proposition \ref{prop:HS}.
%Corollary~\ref{cor:optimalRepairAlg} can be proved by Definition~\ref{def:just} and
%Definition~\ref{def:optimalRepair}.
%We could use it to compute all optimal repairs when the intersection of all justifications is not empty.
%Note that, when the justification core Note, we could not get all optimal repairs.
%
%??? How to compute all optimal repairs if the core is not empty and without using Proposition 1?
%
The application of the union of all justifications can be used as a step towards deducing IAR entailments~\cite{journals/ki/Penaloza20}.  

% !TEX root =  dl21.tex

\section{Evaluation}
\label{sec:evaluation}

To evaluate the performance of our algorithms in real-world ontologies,
we built a prototypical implementation.
The black-box algorithm is implemented in Java and uses
the OWL~API~\cite{HorridgeBechhofer2011} to access ontologies and 
HermiT~\cite{GlimmHorrocks2014} as a standard reasoner. 
The MUS-membership algorithm (MUS-MEM) is implemented in Python and calls cmMUS~\cite{janota2011cmmus} to detect 
whether a clause is a member of MUSes. 
% A set of 95 ontologies are used in the evaluation.
The ontologies used in the evaluation come from the classification task at the ORE competition 2014~\cite{ParsiaMatentzoglu2015}.
Among them, we selected the ontologies that have less than 10,000 axioms, for a total of 95 ontologies.
%In total, we test 95 ontologies in the evaluation.
In the experiments, we computed a single justification, the intersection and union of all justifications 
for all atomic concept inclusions that are entailed by the ontologies.%
\footnote{An atomic concept inclusion is the inclusion that in the form of $A\sqsubseteq B$, where $A$ and $B$ are concept names.} 
All experiments ran on two processors Intel® Xeon® E5-2609v2 2.5GHz, 8 cores, 64Go, Ubuntu 18.04. 
All the figures in this section plot the logarithmic computation time (in the vertical axis) of each test instance (in the horizontal axis).

% In the following, a query refers a axiom $\alpha:A\sqsubseteq B$. We will compute the core and the union of justification of each query over a selected set of ontologies. We say a query $\alpha$ is trivial if the justification $\Just(\Omc, \alpha)$ contain only one element and non-trivial if not. In order to make clear illustration, we use the log value for time cost of each methods in all the figures.

% \begin{figure}[tb]
% \begin{floatrow}
% \ffigbox{%
%   \includegraphics[width=0.5\textwidth]{fig/comparing_core_all.png} 
% }{%
%   \caption{Computation time of \Cmc vs \Jmc}%
%   \label{compare-core}
% }

% \capbtabbox{%
% \centering
%   \begin{tabular}{@{}c@{\quad }r@{\quad }r@{}} 
%   \toprule
%   &\textbf{$\Jmc$}\mbox{\hspace{0.55em}}& \textbf{$\Cmc$}\mbox{\hspace{0.6em}}\\
%       \midrule
%         \textbf{min}&0.001s&0.001s\\
%         \textbf{max} &137.245s&\;260.002s\\
%         \textbf{mean}&0.206s&0.318s\\
%         \textbf{median}&0.005s&0.001s\\
%         \bottomrule
%         %\\[4.5mm]
%   \end{tabular}
%   \vspace{4.5mm}
% }{%
% \caption{Statistics of Fig.~\ref{compare-core}}%
% \label{tab:core_just}
% }
% \end{floatrow}
% %\caption{Computation time of the intersection of all justifications and one justification}
% \end{figure}

\begin{figure}[tb]
\begin{floatrow}
\ffigbox{%
  \includegraphics[width=0.5\textwidth]{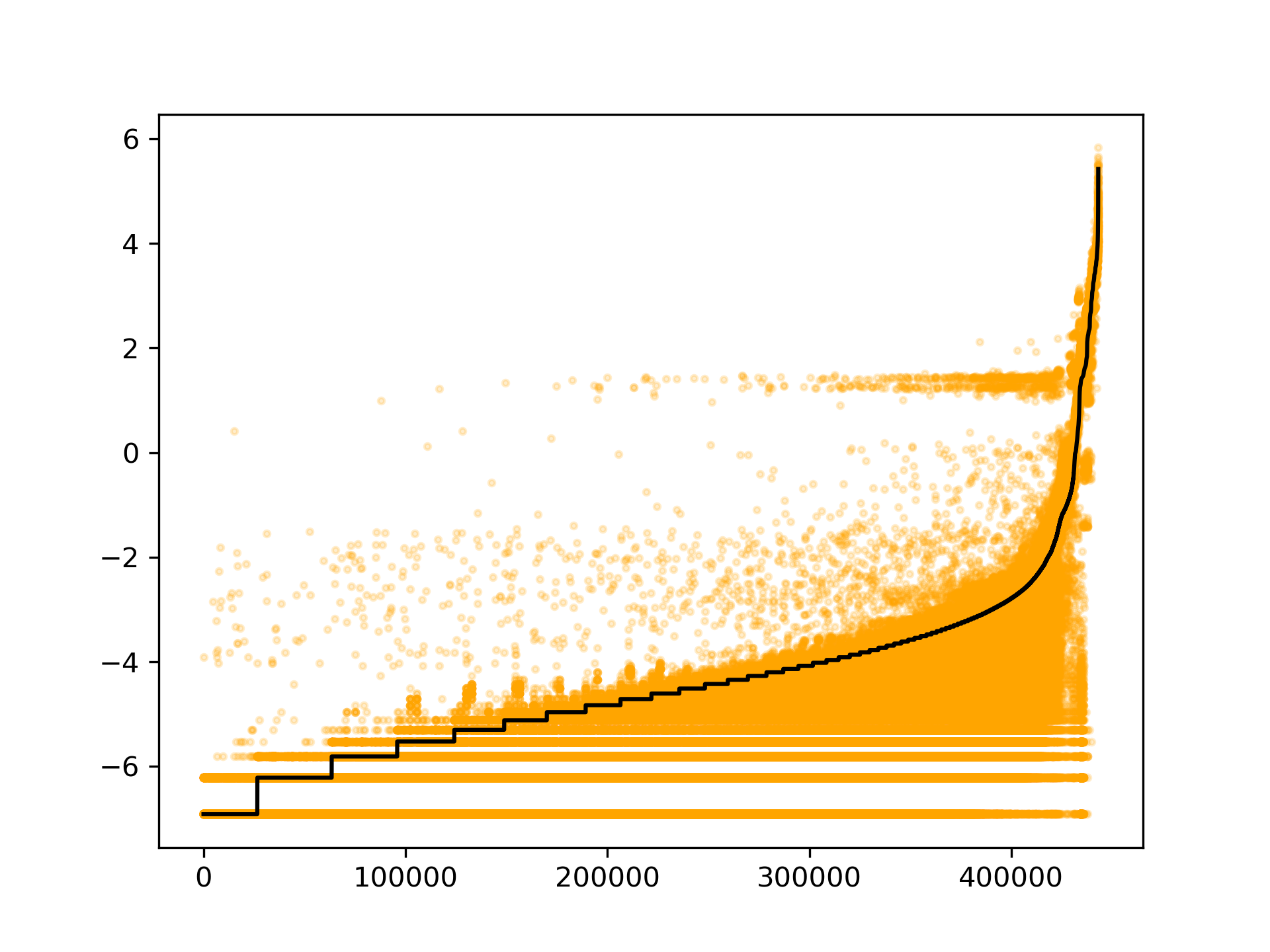} 
}{%
  \caption{Computation time of \Cmc vs \Jmc}%
  \label{compare-core}
}

\capbtabbox{%
\centering
  \begin{tabular}{@{}c@{\quad }r@{\quad }r@{}} 
  \toprule
  &\textbf{$\Jmc$}\mbox{\hspace{0.55em}}& \textbf{$\Cmc$}\mbox{\hspace{0.6em}}\\
      \midrule
        \textbf{min}&0.001s&0.001s\\
        \textbf{max} &226.608s&\;341.560s\\
        \textbf{mean}&0.400s&0.456s\\
        \textbf{median}&0.009s&0.002s\\
        \bottomrule
        %\\[4.5mm]
  \end{tabular}
  \vspace{4.5mm}
}{%
\caption{Statistics of Fig.~\ref{compare-core}}%
\label{tab:core_just}
}
\end{floatrow}
%\caption{Computation time of the intersection of all justifications and one justification}
\end{figure}

\paragraph{\textbf{Computation time of the core vs. a single justification.}}
Fig.~\ref{compare-core} compares the time to compute the core against computing a single justification.
The instances in the horizontal axis are ordered according to the single-justification computation time, represented by the black
line. Orange dots represent the core computation time through Algorithm~\ref{alg:compute_core}.
Table~\ref{tab:core_just} provides some basic statistics for comparison.
Generally, computing the core is almost as fast as computing one justification as expected.
Note that, in terms of computational complexity computing the core and one justification are equally hard problems,
the size of the remaining ontology reduces during the latter process. 
Intuitively, if $\Omc' \subseteq \Omc$, checking whether a subsumption is satisfied by $\Omc'$ would be faster than checking it on 
$\Omc$.

%-how many cases that have only one justification
%Moreover, as shown in Fig. \ref{percent_ratio}, Among all the queries, only 2.38\% queries have empty core and 83.27\% queries is trivial therefore have only one justification. Among the non-trivial queries, only 14.17\% of them have empty core.

\begin{figure}[tb]
\begin{floatrow}
\ffigbox{%
  \includegraphics[width=0.5\textwidth]{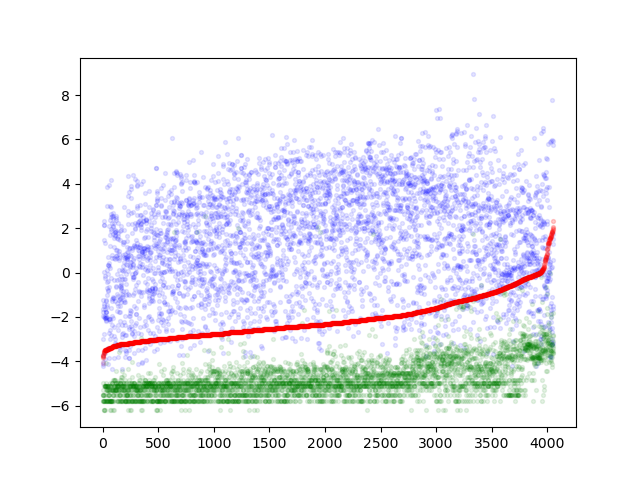} 
}{%
  \caption{Computation time of the union for \ALC-ontologies when there exist several justifications}
        \label{compare-union-non-trivial}
}

\ffigbox{%
  \includegraphics[width=0.5\textwidth]{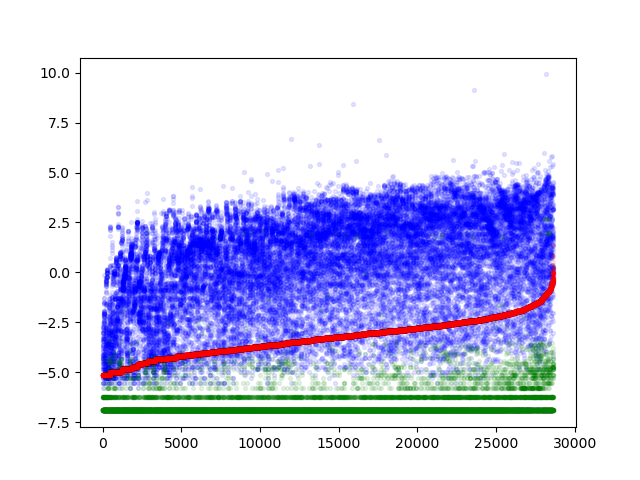} 
}{%
  \caption{Computation time of the union for \ALC-ontologies when there exists one justification}
\label{compare-union-trivial}
}
\end{floatrow}
% \end{figure}

% \begin{figure}
\begin{floatrow}
\ffigbox{%
   \includegraphics[width=0.5\textwidth]{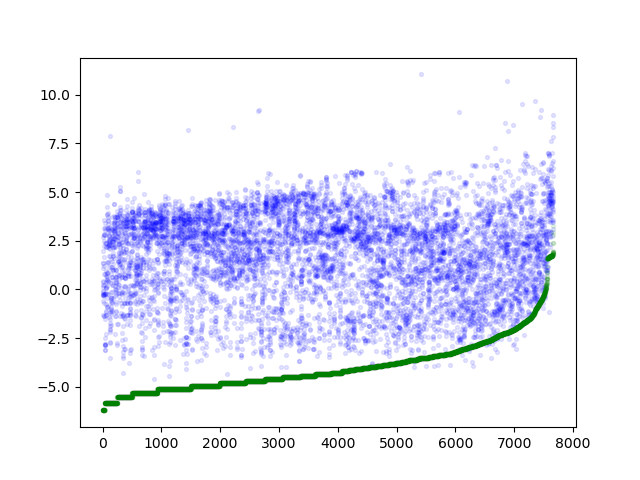}
}{%
 \caption{Computation time of the union for more expressive ontologies when there exist several justifications}
        \label{nonALC-non-trivial}
}

\ffigbox{%
   \includegraphics[width=0.5\textwidth]{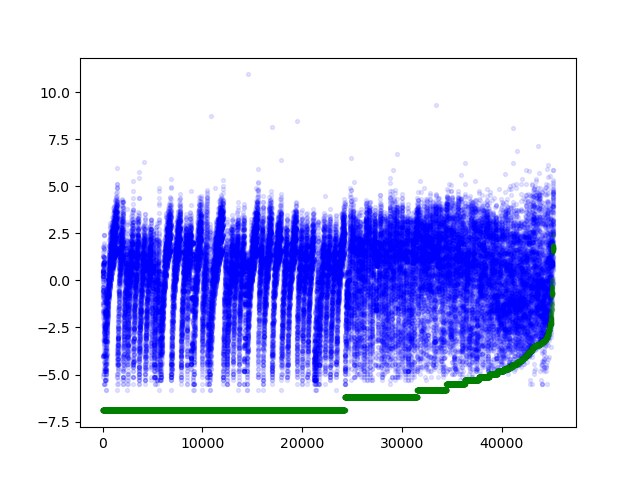} 
}{%
  \caption{Computation time of the union for more expressive ontologies when there exists one justification}
\label{nonALC-trivial}
}

\end{floatrow}
\end{figure}

\paragraph{\textbf{Computation time of the union of all justifications.}}
As a benchmark, we use OWL~API to compute all justifications and then get the union. 
As our second algorithm could compute the union of all justifications only for \ALC-ontologies,  
we separate our ontologies into two categories: one is \ALC-ontologies and the other one is the ontologies that are more expressive 
than \ALC. 
The computation time for the union of all justifications for \ALC-ontologies is shown in Fig.~\ref{compare-union-non-trivial} 
(the cases with several justifications) and Fig.~\ref{compare-union-trivial} (the cases with only one justification).
Figs.~\ref{nonALC-non-trivial} and~\ref{nonALC-trivial} show the computation time of the union of all justifications for more expressive 
ontologies when there exists multiple justifications and only one justification respectively.
In Figs.~\ref{compare-union-non-trivial}--\ref{nonALC-trivial}, 
%Y-axis represents the computation time of the union of all justifications in log scale and X-axis represents each individual conclusion. 
%Furthermore, 
each blue, green or red dot corresponds to computation time of the union by OWL~API, the black-box algorithm or the MUS-MEM 
algorithm for a conclusion respectively. 
We order the conclusions along the X-axis by increasing order of computation time of MUS-MEM algorithms in 
Figs.~\ref{compare-union-non-trivial} and Fig.~\ref{compare-union-trivial}, and by the black-box performance in the latter two
figures.
We observe from these plots that the black-box algorithm outperforms other methods, and when available, MUS-MEM tends to perform
better than a direct use of the OWL~API.

\begin{figure}[tb]
    \centering
   \ffigbox{%
    \includegraphics[width=12.5cm]{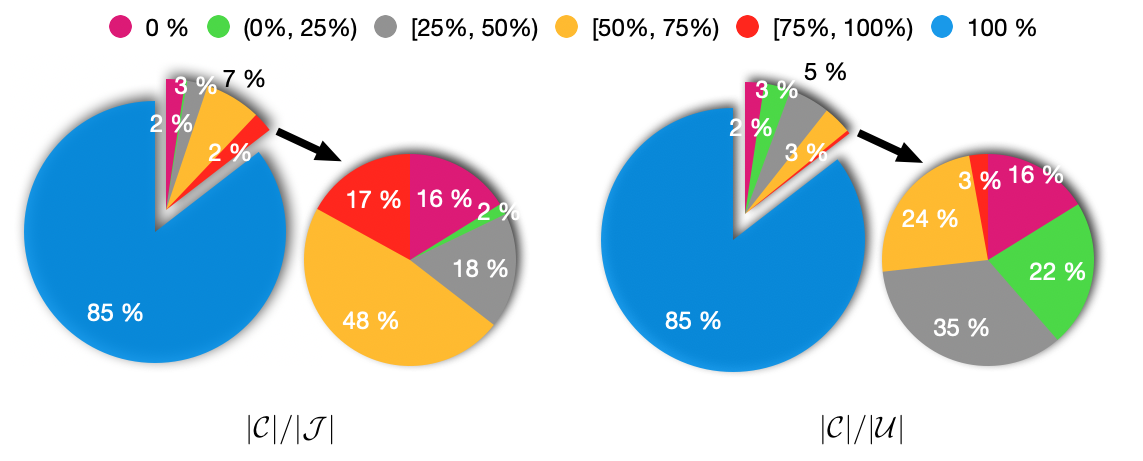}
    }{
    %\caption{Ratio of the size of the intersection of all justifications to the size of one random justification}
    \caption{Ratio of $|\Cmc|$ to a random $|\Jmc|$ (left) and ratio of $|\Cmc|$ to $|\Umc|$ (right).}
     \label{percent_ratio}
    }
\end{figure}

\paragraph{\textbf{Size comparisons for justifications, cores, and unions of justifications.}}
Fig.~\ref{percent_ratio} illustrates the ratio of the size of the core to the size of a random justification and 
to the size of the union of all justifications.
In our experiments, the intersection of all justifications for only 2.35\% subsumptions is empty, which means that we could 
use Corollary~\ref{cor:optimalRepairAlg} to compute optimal repairs for 97.65\% of the cases.
Moreover, for more than 85\% cases, the size of a justification ($|\Jmc|$) equals to the size of the core ($|\Cmc|$), which indicates that 
there exists only one justification.
When several justifications exist (the second chart from the left of Fig.~\ref{percent_ratio}), the ratio of $|\Cmc|$ to a random 
$|\Jmc|$ falls between 50\% to 75\% for almost half of the cases.
The right-most chart displays the distribution of the ratio of $|\Cmc|$ to the union of 
all justifications $\Umc$ when there exist multiple justifications.
The ratio distributes quite evenly between 0\% (not including) to 75\%. 
Interestingly, the intersection of all justifications is empty for only 16\% subsumptions even when several justifications exist.

% !TEX root =  dl21.tex

\section{Conclusions}
\label{sec:conclusion}

In this paper, we presented algorithms for computing the core (that is, the intersection of all justifications) and
the union of all justifications for a given DL consequence. Most of the algorithms are based on repeated calls to a
(black-box) reasoner, and hence apply for ontologies and consequences of any expressivity, as long as a reasoner exists.
The only exception is a MUS-based approach for computing the union of all justifications, which depends on the properties
of the \ALC consequence-based method implemented by \emph{condor}. Still, the approach should be generalisable without
major problems to any language for which consequence-based reasoning methods exists like, for instance, 
$\mathcal{SROIQ}$ \cite{CuGH19,CuGH18}.

As an application of our work, we study how to find optimal repairs effectively, through the information provided by the core
and the union of all justifications. Through an empirical analysis, run over more than 100,000 consequences from almost a
hundred ontologies from the ORE 2014 competition we observe that our methods behave better in practice than the usual
approach through the OWL API. 
A more detailed analysis of the experimental results is left for future work.

Our experiments also confirm an observation that has already been made for light-weight ontologies \cite{Sunt-PhD09}, and
to a smaller degree in the ontologies from the BioPortal corpus \cite{Bail-PhD13}; namely, that consequences tend to have one, 
or only a few, overlapping justifications. In our case, we exploit this fact, and the efficient core computation algorithm to find optimal
repairs in more than 97\% of the test instances: those with exactly one justification, where removing any axioms from it leads to
an optimal repair.

%Moreover, we present their application: how to do optimal repair with the intersection and the union of all justifications.
%Our initial experiment showed that our approaches are much faster than
%state-of-the-art methods.
%%We further present how to use the intersection and the union of all justifications to repair ontologies. 
%Additionally, we can perform optimal repair for more than 97\% cases by using the intersection of all justifications.  

\bibliographystyle{abbrv}
\bibliography{ref}

\begin{thebibliography}{10}

\bibitem{beacon}
M.~F. Arif, C.~Menc{\'\i}a, A.~Ignatiev, N.~Manthey, R.~Pe{\~n}aloza, and
  J.~Marques-Silva.
\newblock Beacon: An efficient sat-based tool for debugging
  \(\mathcal{EL}\)\({}^{\mbox{+}}\) ontologies.
\newblock In {\em International Conference on Theory and Applications of
  Satisfiability Testing}, pages 521--530. Springer, 2016.

\bibitem{EL2MCS}
M.~F. Arif, C.~Menc{\'\i}a, and J.~Marques-Silva.
\newblock Efficient axiom pinpointing with {EL2MCS}.
\newblock In {\em Joint German/Austrian Conference on Artificial Intelligence
  (K{\"u}nstliche Intelligenz)}, pages 225--233. Springer, 2015.

\bibitem{EL2MUS}
M.~F. Arif, C.~Menc{\'\i}a, and J.~Marques-Silva.
\newblock Efficient {MUS} enumeration of {Horn} formulae with applications to
  axiom pinpointing.
\newblock In {\em International Conference on Theory and Applications of
  Satisfiability Testing}, pages 324--342. Springer, 2015.

\bibitem{baader1995embedding}
F.~Baader and B.~Hollunder.
\newblock Embedding defaults into terminological knowledge representation
  formalisms.
\newblock {\em Journal of Automated Reasoning}, 14(1):149--180, 1995.

\bibitem{BaPe-JAR10}
F.~Baader and R.~Pe{\~{n}}aloza.
\newblock Automata-based axiom pinpointing.
\newblock {\em J. Autom. Reason.}, 45(2):91--129, 2010.

\bibitem{BaPe-JLC10}
F.~Baader and R.~Pe{\~{n}}aloza.
\newblock Axiom pinpointing in general tableaux.
\newblock {\em J. Log. Comput.}, 20(1):5--34, 2010.

\bibitem{BaPS-KI07}
F.~Baader, R.~Pe{\~{n}}aloza, and B.~Suntisrivaraporn.
\newblock Pinpointing in the description logic {EL+}.
\newblock In J.~Hertzberg, M.~Beetz, and R.~Englert, editors, {\em Proceedings
  of the 30th Annual German Conference on AI, {KI} 2007}, volume 4667 of {\em
  Lecture Notes in Computer Science}, pages 52--67. Springer, 2007.

\bibitem{Bail-PhD13}
S.~P. Bail.
\newblock {\em The justificatory structure of {OWL} ontologies}.
\newblock PhD thesis, University of Manchester, {UK}, 2013.

\bibitem{condor}
A.~Bate, B.~Motik, B.~C. Grau, D.~T. Cucala, F.~Siman\v{c}\'{\i}k, and
  I.~Horrocks.
\newblock Consequence-based reasoning for description logics with disjunctions
  and number restrictions.
\newblock {\em J. Artif. Int. Res.}, 63(1):625–690, Sept. 2018.

\bibitem{chang2014symbolic}
C.-L. Chang and R.~C.-T. Lee.
\newblock {\em Symbolic logic and mechanical theorem proving}.
\newblock Academic press, 2014.

\bibitem{ChenLMW-ISWC17}
J.~Chen, M.~Ludwig, Y.~Ma, and D.~Walther.
\newblock Zooming in on ontologies: Minimal modules and best excerpts.
\newblock In {\em Proc.\ of {ISWC'17}, Part {I}}, volume 10587 of {\em Lecture
  Notes in Computer Science}, pages 173--189. Springer, 2017.

\bibitem{CuGH18}
D.~T. Cucala, B.~C. Grau, and I.~Horrocks.
\newblock Consequence-based reasoning for description logics with disjunction,
  inverse roles, number restrictions, and nominals.
\newblock In J.~Lang, editor, {\em Proceedings of the Twenty-Seventh
  International Joint Conference on Artificial Intelligence, {IJCAI} 2018},
  pages 1970--1976. ijcai.org, 2018.

\bibitem{CuGH19}
D.~T. Cucala, B.~C. Grau, and I.~Horrocks.
\newblock Sequoia: {A} consequence based reasoner for {SROIQ}.
\newblock In M.~Simkus and G.~E. Weddell, editors, {\em Proceedings of the 32nd
  International Workshop on Description Logics}, volume 2373 of {\em {CEUR}
  Workshop Proceedings}. CEUR-WS.org, 2019.

\bibitem{GlimmHorrocks2014}
B.~Glimm, I.~Horrocks, B.~Motik, G.~Stoilos, and Z.~Wang.
\newblock {HermiT}: an {OWL} 2 reasoner.
\newblock {\em Journal of Automated Reasoning}, 53(3):245--269, 2014.

\bibitem{HorridgeBechhofer2011}
M.~Horridge and S.~Bechhofer.
\newblock The {OWL API}: A {Java} {API} for {OWL} ontologies.
\newblock {\em Semantic Web}, 2(1):11--21, 2011.

\bibitem{janota2011cmmus}
M.~Janota and J.~Marques-Silva.
\newblock cm{MUS}: A tool for circumscription-based mus membership testing.
\newblock In {\em International Conference on Logic Programming and
  Nonmonotonic Reasoning}, pages 266--271. Springer, 2011.

\bibitem{KPHS07}
A.~Kalyanpur, B.~Parsia, M.~Horridge, and E.~Sirin.
\newblock Finding all justifications of {OWL} {DL} entailments.
\newblock In {\em Proceedings of ISWC 2007 \& ASWC 2007}, volume 4825 of {\em
  LNCS}, pages 267--280. Springer, 2007.

\bibitem{kalyanpur2005debugging}
A.~Kalyanpur, B.~Parsia, E.~Sirin, and J.~Hendler.
\newblock Debugging unsatisfiable classes in {OWL} ontologies.
\newblock {\em Journal of Web Semantics}, 3(4):268--293, 2005.

\bibitem{kalyanpur2006debugging}
A.~A. Kalyanpur.
\newblock {\em Debugging and repair of OWL ontologies}.
\newblock PhD thesis, 2006.

\bibitem{PULi}
Y.~Kazakov and P.~Sko{\v{c}}ovsk{\`y}.
\newblock Enumerating justifications using resolution.
\newblock In {\em International Joint Conference on Automated Reasoning}, pages
  609--626. Springer, 2018.

\bibitem{conf/ijcai/KoopmannC20}
P.~Koopmann and J.~Chen.
\newblock Deductive module extraction for expressive description logics.
\newblock In C.~Bessiere, editor, {\em Proceedings of {IJCAI'20}}, pages
  1636--1643. ijcai.org, 2020.

\bibitem{liberatore2005redundancy}
P.~Liberatore.
\newblock Redundancy in logic i: Cnf propositional formulae.
\newblock {\em Artificial Intelligence}, 163(2):203--232, 2005.

\bibitem{LiSa-SAT05}
M.~H. Liffiton and K.~A. Sakallah.
\newblock On finding all minimally unsatisfiable subformulas.
\newblock In F.~Bacchus and T.~Walsh, editors, {\em Proceedings of the 8th
  International Conference on Theory and Applications of Satisfiability Testing
  ({SAT} 2005)}, volume 3569 of {\em Lecture Notes in Computer Science}, pages
  173--186. Springer, 2005.

\bibitem{SATpin}
N.~Manthey, R.~Pe{\~n}aloza, and S.~Rudolph.
\newblock Efficient axiom pinpointing in {EL} using sat technology.
\newblock In {\em Description Logics}, 2016.

\bibitem{ParsiaMatentzoglu2015}
B.~Parsia, N.~Matentzoglu, R.~S. Gon{\c{c}}alves, B.~Glimm, and A.~Steigmiller.
\newblock The {OWL} reasoner evaluation ({ORE}) 2015 competition report.
\newblock {\em Journal of Automated Reasoning}, pages 1--28, 2015.

\bibitem{parsia2005debugging}
B.~Parsia, E.~Sirin, and A.~Kalyanpur.
\newblock Debugging owl ontologies.
\newblock In {\em Proceedings of the 14th international conference on World
  Wide Web}, pages 633--640, 2005.

\bibitem{Pena-PP20}
R.~Pe{\~{n}}aloza.
\newblock Axiom pinpointing.
\newblock In G.~Cota, M.~Daquino, and G.~L. Pozzato, editors, {\em Applications
  and Practices in Ontology Design, Extraction, and Reasoning}, volume~49 of
  {\em Studies on the Semantic Web}, pages 162--177. {IOS} Press, 2020.

\bibitem{journals/ki/Penaloza20}
R.~Pe{\~{n}}aloza.
\newblock Error-tolerance and error management in lightweight description
  logics.
\newblock {\em K{\"{u}}nstliche Intell.}, 34(4):491--500, 2020.

\bibitem{conf/esws/PenalozaMIM17}
R.~Pe{\~{n}}aloza, C.~Menc{\'{\i}}a, A.~Ignatiev, and J.~Marques{-}Silva.
\newblock Lean kernels in description logics.
\newblock In E.~Blomqvist, D.~Maynard, A.~Gangemi, R.~Hoekstra, P.~Hitzler, and
  O.~Hartig, editors, {\em Proceeding of {ESWC'17}}, volume 10249 of {\em
  Lecture Notes in Computer Science}, pages 518--533, 2017.

\bibitem{ReiterDiagnosis}
R.~Reiter.
\newblock A theory of diagnosis from first principles.
\newblock {\em Artificial Intelligence}, 32(1):57--95, 1987.

\bibitem{ScCo-IJCAI03}
S.~Schlobach and R.~Cornet.
\newblock Non-standard reasoning services for the debugging of description
  logic terminologies.
\newblock In G.~Gottlob and T.~Walsh, editors, {\em Proceedings of the
  Eighteenth International Joint Conference on Artificial Intelligence}, pages
  355--362. Morgan Kaufmann, 2003.

\bibitem{SeVe-CADE09}
R.~Sebastiani and M.~Vescovi.
\newblock Axiom pinpointing in lightweight description logics via {Horn-SAT}
  encoding and conflict analysis.
\newblock In R.~A. Schmidt, editor, {\em Proceedings of the 22nd International
  Conference on Automated Deduction}, volume 5663 of {\em Lecture Notes in
  Computer Science}, pages 84--99. Springer, 2009.

\bibitem{Sunt-PhD09}
B.~Suntisrivaraporn.
\newblock {\em Polynomial time reasoning support for design and maintenance of
  large-scale biomedical ontologies}.
\newblock PhD thesis, Dresden University of Technology, Germany, 2009.

\bibitem{10.1007/978-3-540-89704-0_1}
B.~Suntisrivaraporn, G.~Qi, Q.~Ji, and P.~Haase.
\newblock A modularization-based approach to finding all justifications for
  {OWL DL} entailments.
\newblock In J.~Domingue and C.~Anutariya, editors, {\em The Semantic Web},
  pages 1--15, Berlin, Heidelberg, 2008. Springer Berlin Heidelberg.

\end{thebibliography}

\end{document}